\documentclass[aps,pra,reprint]{revtex4-1}

\usepackage{graphicx}
\usepackage{bm}

\newcommand{\unit}[1]{\ensuremath{\, \mathrm{#1}}}
\newcommand{\rmd}{\mathrm{d}}

\begin{document}

\title{Single Photon Emission from Site-Controlled InGaN/GaN Quantum Dots}

\author{Lei Zhang}
\affiliation{Department of Physics, University of Michigan, 450 Church Street, Ann Arbor, MI 48109, USA}
\author{Chu-Hsiang Teng}
\affiliation{Department of Electrical Engineering and Computer Science, University of Michigan, 1301 Beal Ave., Ann Arbor, MI 48109, USA}
\author{Tyler A. Hill}
\affiliation{Department of Physics, University of Michigan, 450 Church Street, Ann Arbor, MI 48109, USA}
\author{Leung-Kway Lee}
\affiliation{Department of Electrical Engineering and Computer Science, University of Michigan, 1301 Beal Ave., Ann Arbor, MI 48109, USA}
\author{Pei-Cheng Ku}
\email{peicheng@umich.edu}
\affiliation{Department of Electrical Engineering and Computer Science, University of Michigan, 1301 Beal Ave., Ann Arbor, MI 48109, USA}
\author{Hui Deng}
\email{dengh@umich.edu}
\affiliation{Department of Physics, University of Michigan, 450 Church Street, Ann Arbor, MI 48109, USA}

\date{\today}

\begin{abstract}
Single photon emission was observed from site-controlled InGaN/GaN quantum dots. The single-photon nature of the emission was verified by the second-order correlation function up to 90~K, the highest temperature to date for site-controlled quantum dots. Micro-photoluminescence study on individual quantum dots showed linearly polarized single exciton emission with a lifetime of a few nanoseconds. The dimensions of these quantum dots were well controlled to the precision of state-of-the-art fabrication technologies, as reflected in the uniformity of their optical properties. The yield of optically active quantum dots was greater than 90\%, among which 13\%-25\% exhibited single photon emission at 10~K.
\end{abstract}

\pacs{0000.0000}

\maketitle

Semiconductor quantum dots (QDs) have diverse quantum photonic applications\cite{Shields2007,Ladd2010} due to their atomic-like properties, characterized by discrete, optically active energy states. Many critical capabilities of QDs have been demonstrated, including single photon emission,\cite{Michler2000} entanglement generation,\cite{Chen2000,Stevenson2006a} strong coupling with optical cavities,\cite{Reithmaier2004, Yoshie2004, Peter2005} single-QD lasing,\cite{Nomura2010} and control of single spin states.\cite{Ladd2010} However, most work to date has been based on self-assembled QDs in III-As and III-P materials, which face severe limitations in operating temperature and scalability. III-As and III-P QDs typically operate at cryogenic temperatures due to the relatively small exciton binding energies and QD-barrier band offsets. Furthermore, self-assembled QDs are formed at random locations and suffer from large inhomogeneity in size and spectral distribution, which prevents controlled coupling of multiple QDs or coupling of QDs with cavities. In this letter, we report single photon emission from site-controlled single InGaN/GaN QDs that are scalable for manufacturing and can be readily integrated with cavities.

To achieve cryo-free operation, III-N QDs with large QD-barrier band offsets and exciton binding energies have emerged as one of the most promising solutions. Single photon emission has been reported up to 200~K, although only in a self-assembled GaN/AlN QD\cite{Kako2006b} and an InGaN QD in a self-organized AlGaN nanowire.\cite{Deshpande2013} To control the position of a III-N QD, various approaches have been explored.\cite{Edwards2004, He2005, Keller2006, Chen2006a, Chen2007, Hsu2011, Choi2013} To date, QD-like emission has only been reported from dots formed at the apex of a site-controlled GaN pyramid\cite{Edwards2004, Hsu2011}  and dots at the top of a site-controlled AlGaN nanowire.\cite{Choi2013} But no single photon emission has been reported from these systems yet.

In this work, we fabricated site-controlled InGaN/GaN QDs by dry etching of a planar single quantum well (QW). This structure has been applied to III-As systems but with very limited success, due to the detrimental surface effects introduced by the etched sidewall.\cite{Forchel1996} In contrast, the surface recombination velocity of III-N is 2-3 orders of magnitude slower than that of III-As.\cite{Mayer1990, Schlager2008} Bright luminescence has been observed from ensembles of III-N nano-pillars etched from multiple QW structures,\cite{He2005, Keller2006, Chen2006a} and from a single quantum disk at the room temperature.\cite{Lee2011} Yet no clear evidence of single photon emission from these structures have been reported so far. Here we not only show single photon emission from our lithographically defined, scalable single InGaN/GaN QDs but also show that this quantum phenomenon survives at temperatures up to 90 K, the highest temperature for all site-controlled QDs to date.

\begin{figure}
\includegraphics{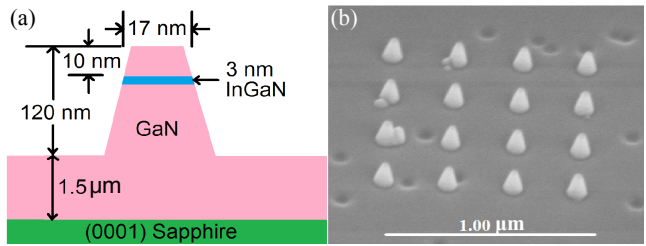}
  \caption{\label{fig:structure} (a) A schematic of a single InGaN QD contained in a GaN nanopillar fabricated by plasma etching of a patterned QW.
  (b) The SEM image of a $1~\mu$m$\times 1~\mu$m dense array of nanopillars of 17~nm in diameter.}
\end{figure}

The sample used for this work was grown by metalorganic chemical vapor deposition (MOCVD) on a sapphire (0001) substrate. A single InGaN QW of 3~nm thickness and 15~\% indium fraction was sandwiched between a 10~nm thick GaN layer on the top and a 1.5~$\mu$m thick GaN layer at the bottom. The planar structure was then patterned via electron beam lithography and subsequently etched into pillars of $\sim120$~nm in height by inductively-coupled plasma reactive-ion etching. No further surface treatment was applied to the sample. The finished device contains multiple arrays of pillars, each pillar, on their top, containing a single InGaN nanodisk as the QD active medium. Details of the fabrication process can be found in our previous work. \cite{Lee2011a} The separation between QDs in each array is also controllable. In the same sample we have both dense arrays with 300 nm inter-dot separation for ensemble photoluminescence (PL) studies and sparse arrays with 5 $\mu$m inter-dot separation to enable the isolation of single QD for micro-photoluminescence ($\mu$-PL) studies. A schematic of a fabricated nanopillar as well as a scanning electron microscope (SEM) image of a dense array containing QDs of $D=17$~nm is shown in Fig.~\ref{fig:structure}(a) and (b), respectively.

We used a confocal microscopy setup to investigate individual QDs' time-integrated and time-resolved PL properties. All data were taken using a frequency-doubled 780~nm Ti:Sapphire laser with 150~fs pulse duration and 80~MHz repetition rate. The second-order correlation function was measured using a Hanbury-Brown-Twiss (HBT) interferometer. Details of the setup are described in the supplementary material. \footnote{\label{suppl}See supplementary material at [URL will be inserted by AIP].}


\begin{figure}
	\includegraphics{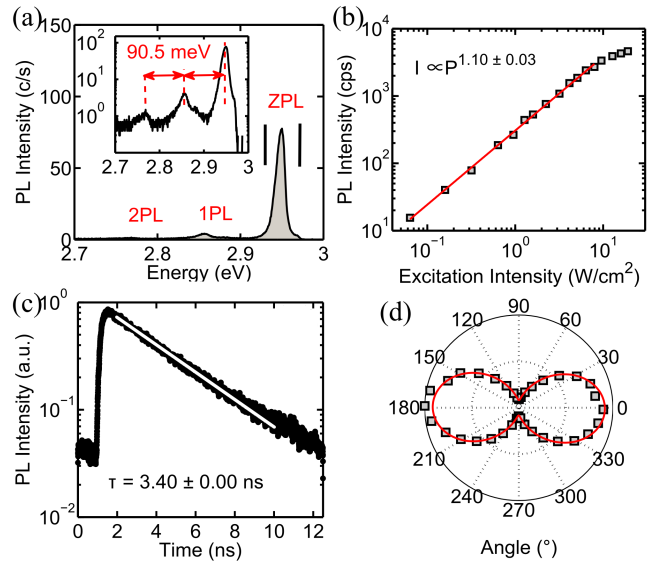}
	\caption{\label{fig:pl10K} 10~K optical properties of a QD of $D = 29$~nm in diameter. (a) The PL spectra at excitation intensity $P = 6.37$~W/cm$^2$. The inset is a semi-logarithm plot of the same spectrum to emphasize the OP replicas. The two black vertical lines represent the spectral window for the $g^{(2)}$ measurement in Fig.~\ref{fig:g2}(a). (b) The PL intensity $I$ vs. the excitation intensity $P$. A fit (solid line) of $\log (I)$ vs. $\log (P)$  shows that $I$ has a linear dependence on $P$. (c) The time-resolved PL decay curve of the ZPL. A mono-exponential fit (white line) shows a 3.40~ns decay time. (d) Exciton emission intensity I vs. angular orientation ($\theta$) of the linear polarizer. The data are fitted with the equation $a cos^2(\theta+b)+c$ (solid line). The fitting results are $a=0.832, b=0^\circ, c=0.08$. The absolute value of the polarizer angle has no physical meaning and is offset so that $b=0^\circ$.}
\end{figure}

The PL spectrum of a single InGaN nanodisk of $\sim 29$~nm in diameter is shown in Fig.~\ref{fig:pl10K}(a), taken at 10~K with an excitation intensity of $P = 6.37$~W/cm$^2$. It consists of a dominant zero-phonon line (ZPL) and its weak optical-phonon (OP) replicas.  The ZPL is at $\sim$2.95~eV with a full-width-at-half-maximum (FWHM) of 14~meV. The energy interval of the series of OP replicas to the lower energy side of the ZPL suggests an OP energy of $\sim$90~meV (inset). The integrated PL intensity $I$ of the ZPL increases linearly with the excitation intensity $P$ as $I \propto P^{1.10}$ (Fig.~\ref{fig:pl10K}(b)). It shows that the ZPL consists of mainly exciton emission. The slight super-linear power dependence may be related to other secondary effects which is unclear at the moment.

The decay of the exciton emission is mono-exponential in time, suggesting that the ZPL consists of mainly single exciton state. The 1/e decay time of 3.4 ns (Fig.~\ref{fig:pl10K}(c)) corresponds to a homogeneous linewidth of $<6$~$\mu$eV, much narrower than the PL linewidth of $\sim$14~meV for the ZPL. This suggests that the ZPL is broadened by fast spectral diffusion\cite{Rice2004a} caused by the charge trapping and releasing processes on the free surfaces nearby, such as the nanodisk sidewall and the top of the capping layer. Therefore the spectral diffusion could be reduced by GaN regrowth after surface treatment. The exciton emission from this QD, as well as from all the QDs we have measured, is linearly polarized as depicted in Fig.~\ref{fig:pl10K}(d). Linearly polarized exciton emission has been observed by several groups in InGaN QDs of different forms.\cite{Hsu2011,Winkelnkemper2007} This has been attributed to the anisotropy of the QD strain profile, possibly caused by the anisotropy in QD geometry.\cite{Winkelnkemper2007} In our QDs, the plasma etching process attacks the sidewall randomly and introduces anisotropy in the lateral geometry, leading to the linear polarization with random orientations.

The above properties of the QD exciton emission remain unchanged at higher excitation densities. Most notably, there is no discernible blueshifts of the exciton energy with the increase of laser-excitation intensity by more than two orders of magnitude. This is in contrast with what is expected from a strained InGaN nanodisk, since it is well known that the strain induces piezo-electric field perpendicular to the InGaN nanodisk plane, which causes Stark shift of the exciton energy. Upon increasing the excitation intensity, more electron-hole pairs (EHPs) are injected into the InGaN layer, which screen the electric field and lead to a blueshift of the exciton energy. Note that, even though the strain is greatly relaxed in our small nanodisks compared to infinitely large planar InGaN QWs,\cite{Hsueh2005, Chen2006a, Kawakami2010, Ramesh2010, Lee2011} a significant amount of strain still remains.\cite{Teng2013} This contradiction can be resolved if our InGaN nanodisk is truly a QD whose excitonic energy levels are energetically separated. In this case the piezo-electric fields will not cause energy blueshifts to QD's spectral lines because of the cascade nature of the QD carrier dynamics, i.e. no matter how many EHPs are injected to begin with, the exciton emission will not occur until all EHPs but one have recombined.

\begin{figure}
	\includegraphics{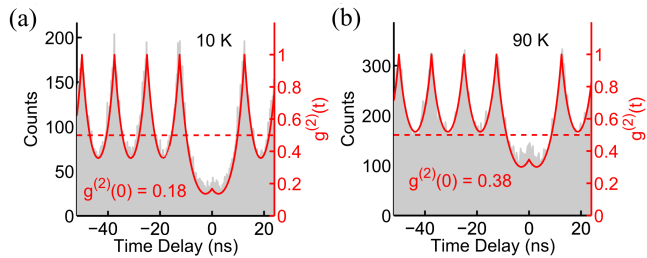}
	\caption{\label{fig:g2} Single photon emission up to 90~K. All data are without any background subtraction. (a) 10~K $g^{(2)}$ data of the QD described in in Fig.~\ref{fig:pl10K}. (b) 90~K $g^{(2)}$ data measured from another QD of $D = 29$~nm in diameter. The solid curves are the fitted $g^{(2)}$ function$^{24}$ showing $g^{(2)}(0)=0.18$ for (a) and $g^{(2)}(0)=0.38$ for (b), respectively.}
\end{figure}

To verify that our InGaN nanodisks are true QDs, capable of generating single photons, we performed second-order correlation ($g^{(2)}$) measurement on the entire ZPL by applying a 30-meV-wide spectral filter as marked by the pair of vertical lines in Fig.~\ref{fig:pl10K}(a). Without saturating the exciton state, at $P = 4.77$~W/cm$^2$ excitation intensity, we obtained the $g^{(2)}$ function of the ZPL as shown in Fig.~\ref{fig:g2}(a). Despite significant overlapping between peaks of neighboring periods due to the short excitation-laser period of $T = 12.5$~ns, the antibunching feature at the zero time delay is evident, with $g^{(2)}(0) \sim 0.22$ without background subtraction. To accurately evaluate the photon-antibunching performance, we fit the pulsed-$g^{(2)}$ data with a model considering the contribution of multiple excitonic states$^{24}$, leading to $g^{(2)}(0) = 0.18$ as shown via the solid curve in Fig.~\ref{fig:g2}(a)), suggesting good single-photon emission from our QDs.

The non-zero $g^{(2)}(0)$ value comes from emission by multi-exciton states within our spectral window. Single photon sources based on QDs typically require a narrow spectra filter to separate the exciton emission from multi-exciton emissions. Due to the large built-in electric field, the emission lines in III-N QDs, including in our QDs, are often severely broadened by spectral diffusion and are difficult to separate, which typically would forbid anti-bunching in the emission. The good anti-bunching in the $g^{(2)}$ data, despite the inclusion of multi-exciton emissions, indicates that excitons have much higher quantum efficiencies (QEs) than multi-exciton states. QE is defined as $\eta=\gamma_{\mathrm{r}}/\gamma$, where $\gamma_{\mathrm{r}}$ is the radiative decay rate and $\gamma$ is the total decay rate. The difference in QEs results from uneven strain-relaxation in the QD. Strain is fully relaxed at the sidewall of the InGaN QD but only partially relaxed at the center, leading to an uneven electric field and a lateral potential along the QD radial direction.\cite{Zhang2013a} The lateral potential is lower at the center and higher near the sidewall; hence it serves as a barrier to protect EHPs from reaching the sidewall surface where they recombine nonradiatively.\cite{Zhang2013c} Therefore, higher potential barriers result in higher EHP QEs. Multi-exciton states screen the electric fields more, yielding lower potential barriers and, consequently, lower QEs. Note that, at low excitation intensities, $g^{(2)}(0)$ is approximately the biexciton-to-exciton QE ratio.\cite{Nair2011a}

Single photon emission was observed up to 90~K from our QDs. With the increase of the temperature, the total PL intensity decreases due to increased thermal escape of the confined EHPs from the lateral confinement potential. Therefore, to study the high temperature properties, we chose another QD with higher PL intensity at 10~K than the previous one. Single-photon emission is demonstrated at 90~K from this QD through the $g^{(2)}$ measurement which shows $g^{(2)} (0)\sim 0.38$ (Fig.~\ref{fig:g2}(b)). This is the highest temperature at which single-photon emission has been observed in site-controlled QDs so far. Interestingly, at 10~K, $g^{(2)}(0)$ of this QD increased to higher than 0.5. This is because, with decreasing temperature, the biexciton's QE is improved, more than that of the exciton's.\cite{Zhang2013c} We would like to note that QD-like PL has been observed up to room temperature in our QDs.\cite{Lee2011} Note also that for InGaN QDs with similar material compositions, high temperature single-photon emission has been reported for a self-assembled InGaN/GaN QD up to 50~K. \cite{Kremling2012} This suggests that even though our fabrication approach involves free surfaces, the optical quality is at least comparable to the highest quality self-assembled InGaN QDs with similar QD-barrier band offsets.

The spectral and statistical properties of our QDs are comparable to those of self-assembled QDs.  \cite{Kremling2012} However, unlike self-assembled QDs, we have control over not only the positioning but also all key structural parameters of our InGaN QDs, including the disk thickness $l$, the disk diameter $D$, and the indium mole fraction $x$, limited only by the present MOCVD growth and electron-beam lithography technologies, as will be discussed below.

\begin{figure}
	\includegraphics{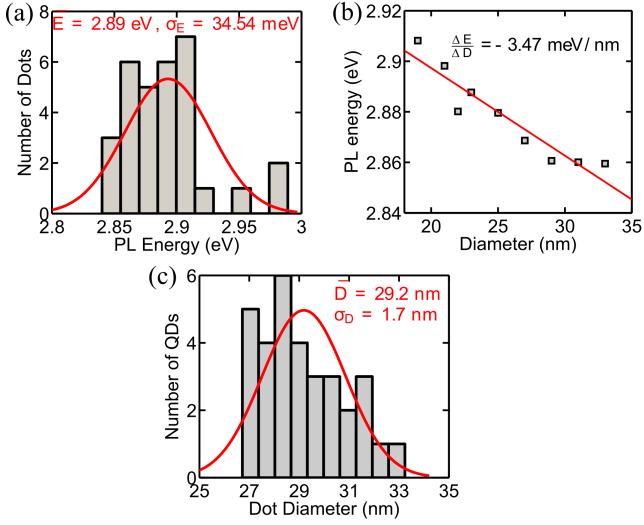}
	\caption{\label{fig:statistics} Uniformity of QD PL energy at 10~K. (a) PL energy distributions of 30 QDs of diameter $D = 29$~nm within 1.7~nm of fluctuation as shown in (c). (b) PL energy $E$ vs. QD diameter $D$ of nine dense arrays of QDs with diameters ranging from 19~nm to 33~nm. The solid line is a linear fit of the data yielding $\frac{\Delta E}{\Delta D} = - 3.5$~meV/nm. (c) Diameter statistics of 30 QDs in the same array measured by SEM showing a standard deviation of $\sim$2~nm.}
\end{figure}

To verify the control over our QDs' structural parameters, we compare the variations in the QDs' optical properties with uncertainties in the structural parameters in the fabrication. On one hand, the statistical distributions of single-dot PL energy are obtained from the $\mu$-PL of 30 individual QDs. As shown in Fig.~\ref{fig:statistics}(a), the PL energy has a standard deviation of 34.5~meV. On the other hand, we can estimate the variation in the QDs' PL energy due to fluctuations in $l$, $D$, and $x$. The results are summarized in Table~\ref{tab:Delta_E}.

\begin{table}
\caption{\label{tab:Delta_E} The contributions of the thickness $l$, indium fraction $x$ and diameter $D$ fluctuations to the total PL energy $E$ inhomogeneity $\Delta E$ for a circular In$_{x=0.15}$Ga$_{0.85}$N nanodisk with $l=3$~nm and $D=29$~nm. $\Delta E$ is calculated from each contribution as $\Delta E = \sqrt{\Delta E_l^2 + \Delta E_x^2 + \Delta E_D^2}$.}
  \begin{tabular}{ l  l  r }
  \hline \hline
  Source              & Fluctuation & Contribution to $\Delta E$ \\
  \hline
  Thickness       & $\Delta l=2$~ML      & $\Delta E_l=28$~meV        \\
  Diameter       &  $\Delta D = 1.7$~nm & $\Delta E_D = 6$~meV       \\
  Indium fraction & $\Delta x = 0.2\%$   & $\Delta E_x = 8$~meV       \\
	\hline	
	\multicolumn{2}{r}{Total calculated}& $\Delta E = 30$~meV      \\
	\hline
	\multicolumn{2}{r}{Total observed}& $\sigma_{E} = 34.5$~meV      \\
	\hline \hline
  \end{tabular}
\end{table}

To evaluate the contribution of the fluctuation of $\Delta l = 2$~ML (1~ML$\sim$0.5~nm) we use a simple planar capacitor model in which the upper and lower InGaN/GaN interfaces are the two capacitor plates. We assume that a fixed amount of charge at the two plates leads to a fixed electric field strength $F$. Therefore, increasing $l$ decreases $E$ by $\Delta E = e F \Delta l$. The amount of $E$ reduction from the intrinsic InGaN bandgap without any strain $E_0$ is $E_0 - E = e F l = 63$~meV which is extracted from the $E$ vs. $D$ relation of a QD.\cite{Zhang2013c} Knowing $l=3$~nm and $\Delta l=2$~ML we get $\Delta E = e F \Delta l / l= 21$~meV. Increasing $l$ also reduces carrier confinement in the vertical (growth) direction which further reduces the PL energy. This can be estimated by a simple 1D particle-in-a-box model, assuming the potential in the vertical direction acts as an infinite potential well, as $\Delta E = \Delta l  \frac{\rmd}{\rmd l} \frac{(\hbar \pi/l)^2}{2m} = 18$~meV in which $\hbar$ is the reduced Planck constant and $m$ is the effective mass of a GaN exciton, 1.6 times the free electron mass. Therefore, the total contribution due to $\Delta l$ is roughly $\Delta E_l = \sqrt{21^2+18^2} = 28$~meV.

To study how the diameter influences the optical properties of our QDs, we measured the PL energy of nine dense arrays of QDs with different diameters. Each dense array has $100 \times 100$ QDs with the same nominal diameter. Fig.~\ref{fig:statistics}(b) shows the PL energy $E$ vs. QD diameter $D$ data, which reflects mainly the relaxation of strain due to diameter reduction.\footnote{Note that, with a reduction of the disk diameter, the PL energy may also increase due to lateral quantum confinement. However, this only leads to a change of $<10$~meV varying the diameter from 30~nm to 20~nm.} A linear fit of the data suggests that the PL energy changes with diameter at a rate $\frac{\Delta E}{\Delta D} = - 3.5$~meV/nm. On the other hand, a statistics of QD diameter in the same array measured by SEM suggests that the diameter fluctuation is $\Delta D = 2$~nm as shown in Fig.~\ref{fig:statistics}(c). Therefore, PL energy inhomogeneity $\Delta E$ caused by diameter fluctuation is only around $\Delta E_D = 6$~meV.

The indium-fraction inhomogeneity originates from the fundamental uncertainty in the number of indium atoms in each QD. Studies \cite{Humphreys2007} have shown that InGaN is a random alloy in which indium atoms distribute randomly but uniformly, and that they do not form the so-called indium-rich islands of the nanometer-length scale especially when the indium fraction is as low as $x=15~\%$ as in our case. In a disk-shaped In$_{0.15}$Ga$_{0.85}$N QD of $3$ nm in thickness and $29$~nm in diameter, there are around $N = 4,500$ indium atoms. Being a random alloy means that the number of indium atoms obeys the Poisson distribution whose standard deviation can be calculated as $\sigma_{N} = \sqrt{N} = 67$. Therefore, the minimum indium-fraction fluctuation is $\Delta x= x \frac{\sigma_{N}}{N} = 0.2\%$. This gives rise to a PL energy fluctuation of $\Delta E_x = (3.5~\unit{eV} - 2.89~\unit{eV})\frac{\Delta x}{x} = 8~\unit{meV}$, in which $3.5$ eV is GaN band gap and $2.89$~eV is the average QD PL energy from Fig.~\ref{fig:statistics}(a). Here we assumed that the band gap varies linearly with the indium fraction.

All three sources of inhomogeneities, as summarized in Table~\ref{tab:Delta_E}, add up to $\Delta E = \sqrt{\Delta E_l^2 + \Delta E_x^2 + \Delta E_D^2} = 30$~meV, which matches well with the observed total PL energy fluctuation of $\sigma_{E} = 34.5$~meV. From this analysis we also concluded that the QD thickness fluctuation is the dominant source of inhomogeneity.

The yield of optically active nanodisks is over 90\% for disks $>20$~nm in diameter. Interestingly, none of the nanodisks smaller than $15$~nm is optically active. Such sharp diameter cut-off provides strong evidence that the formation of our QDs is not due to the random inclusion of so-called localization centers (LCs). Such LCs, expected to be a few nanometers large in the lateral dimension, \cite{Humphreys2007} could form due to QW thickness fluctuations or at indium-rich islands. The sharp intensity cut-off at $\sim 15$~nm clearly suggests that the emission intensity is correlated with the size of the entire InGaN disk rather than small few-nanometer-scale LCs within the InGaN disk. As discussed before, the difference in the strain at the center and sidewall of the nanodisks creates an in-plane confinement potential in large nanodisks, which protects excitons from surface recombination. \cite{Zhang2013c} The depth of the confinement potential decreases as the disk diameter reduces. In disks smaller than 15~nm, the strain relaxation is significant enough even at the center of the disk, and the in-plane confinement potential effectively diminishes, resulting in very low quantum efficiency in these dots.

To estimate the yield of good single photon source from our QDs, we measured $g^{(2)}(0)$ for 16 out of 30 QDs chosen to cover the full range of PL energies. All of them demonstrated photon antibunching ($g^{(2)}(0)<1$) at 10~K while four have $g^{(2)}(0)<0.5$, showing that the yield of single photon source is 13\%-25\%. The variation in $g^{(2)}(0)$ reflects the variation in the biexciton-to-exciton QE ratios, which is again caused by the variation in the height of the confinement potential.\cite{Zhang2013c} QDs with higher confinement potentials have higher biexciton-to-exciton QE ratios and, hence, worse anti-bunching.

In conclusion, we have demonstrated single photon emission from site-controlled single InGaN/GaN QDs up to 90~K, the highest temperature among all site-controlled QDs to date. The structural parameters of the QDs, and correspondingly their optical properties, are well controlled, limited mainly by the precision of the electron-beam lithography and MOCVD technologies. Future improvements, including GaN regrowth and using AlGaN as the barrier material, may reduce the spectral linewidth and increase the QD operating temperature.

These QDs have the potential to serve as cryo-free single-photon sources at wavelengths that match the peak-efficiency wavelength of silicon photo-detectors. They are fabricated in a practical and scalable fashion and can be placed very close to each other, well within a wavelength. Therefore they are suitable for on-chip integration with other nano-photonic components, such as photonic crystal cavities, plasmonic resonators and multiple QDs.

We acknowledge financial supports from the National Science Foundation (NSF) under Awards ECCS 0901477 for the work related to materials properties and device design, ECCS 1102127 for carrier dynamics and related time-resolved measurements, and DMR 1120923 (MRSEC) for work related to light-matter interactions. The work related to epitaxial growth, fabrication, and photon antibunching properties were also partially supported by the Defense Advanced Research Project Agency (DARPA) under grant N66001-10-1-4042. Part of the fabrication work was performed in the Lurie Nanofabrication Facility (LNF), which is part of the NSF NNIN network.


%

\end{document}